\documentstyle[multicol,pra,aps,epsfig]{revtex}

\begin{document}
\draft
\title{Sampling functions for multimode homodyne tomography 
with a single local oscillator}
\author{Jarom\'{\i}r Fiur\'{a}\v{s}ek}
\address{Department of Chemical Physics, The Weizmann Institute of Science, 
Rehovot 76100, Israel}
\address{and Department of Optics, Palack\'{y} University, 17. listopadu 50, 
772 07 Olomouc, Czech Republic}
\date{\today}
\maketitle

\begin{abstract} 

We derive various sampling functions for
 multimode homodyne tomography with a single local oscillator.
These functions allow us to sample multimode  $s$-parametrized 
quasidistributions, 
density matrix elements  in Fock basis, and $s$-ordered moments of 
arbitrary order directly from the measured quadrature statistics.
The  inevitable experimental losses can be compensated by proper
modification  of the sampling functions.  Results of Monte Carlo simulations
for squeezed three-mode state are reported and the feasibility
of reconstruction of the three-mode $Q$-function 
and $s$-ordered moments from $ 10^7$ sampled data
is demonstrated.
\end{abstract}


\begin{multicols}{2}

\section{Introduction}

Recent development of quantum-state reconstruction  methods  
has made it possible to completely reconstruct an unknown  state
of a quantum mechanical system 
provided that many identical copies of the state are available. 
The method was pioneered in quantum optics,  
where an optical homodyne tomography was devised to reconstruct 
a quantum state of traveling electromagnetic field 
\cite{Smithey93a,Breitenbach95,Schiller96,Vasilyev00}.
Other proposed techniques involved unbalanced homodyning \cite{Banaszek96}
and cavity-field measurements by atomic probes \cite{Cavity}.
Quantum-state reconstruction procedures were also successfully applied  
to molecular vibrational state \cite{Dunn95} 
and motional quantum state of trapped ion \cite{Leibfried96}.

Single-mode optical homodyne tomography is now a well-established
technique. Based on balanced homodyne detection, the method seeks to reconstruct 
the quantum state  from the  statistics 
of the quadrature components of the signal mode.
The standard experimental setup involves balanced lossless beam splitter, 
where the signal is mixed with a strong coherent local oscillator (LO).
Two photodetectors are placed at the beam splitter outputs and 
the two photocurrents are subtracted, thereby removing LO fluctuations 
from the resulting signal.

 Wigner function can be obtained
from the measured quadrature statistics by means of inverse Radon 
transform  \cite{Vogel89,Smithey93a}.
Once the Wigner function is known, expectation value of any operator can be
be evaluated by averaging corresponding phase-space function over the Wigner 
quasidistribution. This strategy, however, is not optimal,
because the experimental errors are amplified  during numerical 
data processing and the final error can be very large.
Fortunately, the detour via Wigner function can be avoided 
and density matrix elements in Fock basis can be directly reconstructed 
by averaging appropriate sampling functions over the measured quadrature
 statistics \cite{DAriano94,DAriano95,Leonhardt95,Richter96b,%
Leonhardt96,Richter00}.  
The  functions for sampling $s$-ordered moments were found
in \cite{Richter96,Wunche96,Richter99a}, and those allowing
direct reconstruction of the 
exponential moments of quantum phase distributions were obtained in 
\cite{Dakna98,Fiurasek00} (for a review, see \cite{Welsch99,LeonhardtBook}). 
The problems with inverse Radon transform can be avoided 
by reconstructing smoothed Wigner functions \cite{Richter99}.
The sampling is a simple and straightforward linear operation which
can be in principle performed in real time during experiment. 
We note that, besides linear sampling procedures, reconstruction
strategies based on maximum likelihood estimation \cite{Hradil95}
and maximum entropy principle \cite{Buzek96} have been proposed.

Recently, increasing attention has been devoted to multimode homodyne 
tomography 
\cite{Kuhn95,Raymer96,Richter97,Opatrny97a,McAlister97,McAlister97b,DAriano99}, 
because some of the most interesting quantum mechanical phenomena
stem from correlations between several degrees of freedom.
Let us mention just the EPR paradox and violation of Bell's inequalities 
\cite{Perbook}. Quantized electromagnetic field is one of the most suitable
systems for thorough investigation and exploitation of these phenomena. 
For example, entangled signal and idler photons can be routinely prepared 
by means of spontaneous parametric down-conversion \cite{Kwiat99}.
The entangled photon pairs play crucial role in certain quantum state teleportation 
schemes \cite{Teleportation} and quantum cryptography setups \cite{Jennewein00}.

Multimode extension of optical homodyne tomography is
straightforward.
One can introduce a separate LO and homodyne detector 
for each mode of interest and measure joint multimode quadrature distribution. 
In this case the sampling functions developed
for single-mode tomography can  be immediately employed. 
Very recently, this approach has been used to measure the joint photon 
number statistics of two-mode squeezed state prepared in 
a nondegenerate optical  parametric amplifier \cite{Vasilyev00}.

However, the requirement of specific homodyne detector for each mode 
complicates the experiment. It would often be much more feasible to 
use only one homodyne detector and one LO.
In such experiment, a distribution of one quadrature $X$,
which  is a linear superposition of $N$ single-mode quadratures, is measured. 
The knowledge of the probability distribution of all distinct quadratures $X$
provides a complete information on the multimode quantum state. In particular, 
two-mode tomography with single homodyne detector was discussed extensively. 
The functions for sampling density matrix elements in Fock basis were found in
\cite{Raymer96,Richter97,McAlister97b}, and those allowing 
direct reconstruction of two-mode correlation functions were obtained in
\cite{Opatrny97a,McAlister97,McAlister97b} .
Recently, a general multimode homodyne tomography 
with a single LO was considered and the sampling functions for
density matrix elements were expressed in terms of integrals \cite{DAriano99}.  
In this paper we shall derive various important sampling functions 
for multimode homodyne tomography with a single LO. 
All functions are expressed in analytical form. 
Imperfect detection is considered and it is shown 
that the losses can be compensated by proper modification 
(rescaling) of the sampling functions.

\begin{figure}
\vspace*{6mm}
\centerline{\epsfig{figure=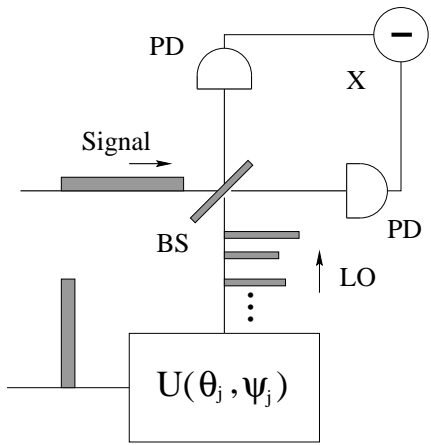,width=0.65\linewidth}}
\vspace*{3mm}
\caption{
Measurement of internal quantum correlations
of optical pulses [29]. The signal pulse and a train of
strong local-oscillator (LO) pulses that are short compared to the 
signal pulse are mixed at a 50\%:50\% beam splitter (BS) and
the two photocurrents measured by photodetectors (PD) are subtracted.
The train of LO pulses is prepared interferometrically, thereby allowing 
one to control the pulse distances, relative phases and intensities,
as symbolically denoted by $U(\theta_j,\psi_j)$.
}
\end{figure}

\begin{figure}
\centerline{\epsfig{figure=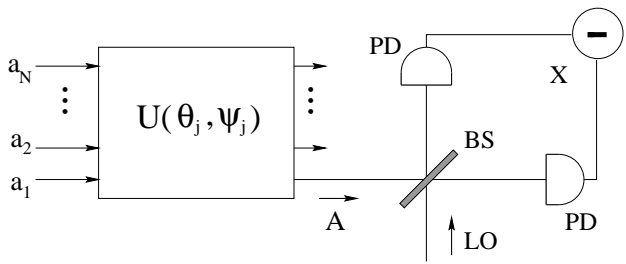,width=0.95\linewidth}}
\vspace*{3mm}
\caption{
Optical homodyne tomography of single-frequency multimode optical field.
The signal modes $a_1,\ldots,a_N$ feed the input of N-port 
interferometr which prepares the mode $A$ at one of its outputs.
Subsequently, the distribution of quadrature $X$ is measured by 
means of standard single-mode homodyne detection.}
\end{figure}

The paper is organized as follows.
In Sec. II we address the reconstruction of multimode smoothed Wigner
functions. The results are then applied in Sec. III to find the 
sampling functions for density matrix elements in Fock basis.   
The reconstruction of $s$-ordered moments of the field operators
is discussed in Sec. IV. The results of Monte Carlo  simulations 
of multimode homodyne tomography  are reported  in Sec. V.
Finally, Section VI contains conclusions.

\section{Sampling functions for $s$-parametrized quasidistributions }

In multimode homodyne tomography with a single local oscillator
one measures a probability distribution of the quadrature
\begin{equation}
X=\frac{1}{\sqrt{2}}(A+A^\dagger),
\end{equation}
where the operator $A$ is a linear superposition of annihilation 
operators $a_j$ of $N$ signal modes,
\begin{equation}
A=\sum_{j=1}^N z_j a_j.
\label{ASUP}
\end{equation}
The complex coefficients $z_l$ fulfill  normalization condition 
\begin{equation}
\sum_{j=1}^N |z_j|^2=1,
\label{ZNORM}
\end{equation}
which ensures validity of standard commutation relation $[A,A^\dagger]=1$
for the operator $A$. Two examples of experimental setups, where the statistics
of quadrature $X$ are measured, are given in Figs. 1 and 2.
Multimode homodyne tomography can be employed to
investigate ultrafast internal quantum correlations of optical pulses
\cite{Opatrny97a,McAlister97}, see Fig. 1. 
A train of N strong LO pulses is used to select a set of
$N$ nonmonochromatic modes from the signal pulse. The modes $a_j$ are
determined by positions and shapes of the LO pulses
and the correlations of the 
signal pulse are probed in terms of these modes.
Figure 2 illustrates a scheme for homodyne tomography of single-frequency 
multimode field. The desired superposition $A$ is prepared in $N$-port
interferometr and then it enters homodyne detector. 
This setup can be used e.g. for measurement of a polarization state of 
an optical field \cite{Raymer99}. The modes $a_1$ and $a_2$ then correspond
to two orthogonal linear polarizations. The two-mode 
unitary transformations $U(\theta,\psi)$ leading to  superpositions
(\ref{ASUP}) can be performed with the help of two phase shifters 
and a polarizing beam splitter \cite{Raymer99}.
A common feature of the experimental setups shown in Figs. 1 and 2 is 
that only one balanced homodyne detector is needed.

If the statistics $w(X;\{z_j\})$ of the quadrature $X$ are known 
for all $\{z_j\}$ fulfilling (\ref{ZNORM}), then we have a complete
knowledge of the quantum state of the multimode light field and all 
quantities of interest, such as various quasidistributions,
density matrix elements,  and  $s$-ordered moments,
can be unambiguously determined from the distributions $w(X;\{z_j\})$.

\subsection{Sampling of the smoothed Wigner functions}

Let us begin with  reconstruction of the multimode
$s$-parametrized quasidistributions.
It is convenient to work in the hyperspherical coordinates. The points
$\{z_j\}$ lie on a surface of $2N$-dimensional unit sphere and we
parametrize them as \cite{DAriano99}
\begin{equation}
z_j=u_j(\bbox{\theta})e^{-i\psi_j},
\label{ZJPARAMETR}
\end{equation}
where
\begin{eqnarray}
u_j(\bbox{\theta})&=&\cos\theta_{j}\prod_{l=1}^{j-1}\sin\theta_l, 
\qquad j<N \\
u_N(\bbox{\theta})&=&\prod_{l=1}^{N-1}\sin\theta_l,
\end{eqnarray}
and
\begin{eqnarray*}
\psi_j\in[0,2\pi],& & \qquad j=1,\ldots,N, \\
\theta_j\in[0,\pi/2],& & \qquad j=1,\ldots,N-1.
\end{eqnarray*}
To simplify the notation, we define 
$\bbox{\theta} = ( \theta_1, \ldots, \theta_{N-1})$ and
$\bbox{\psi}=(\psi_1,\ldots,\psi_N)$.
 
Multimode characteristic function corresponding to $s$-ordering 
of the field operators is  defined as \cite{Perina},
\begin{equation}
C_s(\{\beta_j\})=\left\langle
\prod_{j=1}^N \exp\left(\frac{1}{2}s|\beta_j|^2+\beta_j a_j^\dagger
-\beta_j^\ast a_j\right)\right\rangle,
\label{CDEF}
\end{equation}
where $\langle\rangle$ denotes quantum mechanical average.
Let us compare the exponent on the right-hand side of Eq. (\ref{CDEF})  
with the quadrature  $X(\{z_j\})\equiv X(\bbox{\theta},\bbox{\psi})$.
We can see that $C_s(\{\beta_j\})$ is proportional to  characteristic 
function of the quadrature distribution,
\begin{eqnarray}
C_s\left(\left\{\beta_j\right\}\right)=
e^{sr^2/2}
\int_{-\infty}^{\infty}dX\,e^{i\sqrt{2}rX}w(X;\bbox{\theta},\bbox{\psi}),
\label{CS}
\end{eqnarray}
where $\beta_j=iru_j(\bbox{\theta})\exp(i\psi_j)$ and $r>0$ is radial variable, 
\[
r^2=\sum_{j=1}^N |\beta_j|^2.
\]

Multimode quasidistribution $W_s(\{\alpha_j\})$  is a
Fourier transform of the characteristic function $C_s(\{\beta_j\})$,
\begin{equation}
W_s(\{\alpha_j\})=\frac{1}{\pi^{2N}}
\int C_s(\{\beta_j\})\prod_{j=1}^N d^2\beta_j \,
e^{\beta_j^\ast \alpha_j-\beta_j \alpha_j^\ast}.
\label{WS}
\end{equation}
We rewrite this integral in the hyperspherical coordinates.
We shall integrate
over the angles $\theta_j$, phases $\psi_j$, and
 radius $r$. It is convenient to introduce $d\Omega$,
\begin{equation}
d\Omega= g(\bbox{\theta})\prod_{l=1}^{N-1} d\theta_l \prod _{j=1}^{N} d\psi_j,
\end{equation}
where the prefactor
\begin{equation}
g(\bbox{\theta})=\prod_{l=1}^{N-1}\cos\theta_l (\sin \theta_l)^{2(N-l)-1}
\end{equation}
stems from the Jacobian of coordinate transformation.
We  substitute the  characteristic function (\ref{CS})
into (\ref{WS}) and  after some algebra we arrive at
\begin{eqnarray}
W_s(\{\alpha_j\})&=&
\frac{1}{\pi^{2N}}
\int _{0}^{\infty} \, dr
\int_{\Omega} \, d\Omega
\int_{-\infty}^{\infty} \, dX \,
r^{2N-1}
\nonumber \\
&& \times
e^{sr^2/2}e^{i\sqrt{2}r(X-\tilde{X})}
w(X;\bbox{\theta},\bbox{\psi}),
\label{WSINT}
\end{eqnarray}
where we have introduced a c-number quadrature
\begin{equation}
\tilde{X}(\{\alpha_j\},\bbox{\theta},\bbox{\psi})=
\frac{1}{\sqrt{2}}\sum_{j=1}^N
\left(u_j(\bbox{\theta})e^{-i\psi_j}\alpha_j+{\rm c.c.}\right).
\end{equation}
After changing the order of integration in (\ref{WSINT}), we find that
\begin{eqnarray}
W_s(\{\alpha_j\})&=&
\int_{\Omega}\, d \Omega \int_{-\infty}^{\infty}\, dX \,
w(X;\bbox{\theta},\bbox{\psi})
\nonumber \\ &&\times 
S_N\left(X-\tilde{X}(\{\alpha_j\},\bbox{\theta},\bbox{\psi});s\right),
\label{WSSAMPLED}
\end{eqnarray}
where the sampling function $S_N$ reads
\begin{equation}
S_N\left(\xi;s\right)=
\frac{1}{\pi^{2N}}\int_0^{\infty} d r\, e^{s r^2/2}
e^{i\sqrt{2}r\xi }
r^{2N-1},\qquad s<0.
\label{SNINT}
\end{equation}
This expression can be further simplified. The quasidistributions 
$W_s(\{\alpha_j\})$
as well as  the quadrature distributions $w(X;\bbox{\theta},\bbox{\psi})$ 
are real functions. 
The sampling function $S_N$ has to be  real and only real part of the
above integral should be considered. The imaginary part
of $S_N$ is a null function whose average over any physical
quadrature distribution $w(X;\bbox{\theta},\bbox{\psi})$ is zero.
Thus we can replace 
$\exp(i\sqrt{2}r\xi)$ by $\cos(\sqrt{2}r\xi)$ in Eq. (\ref{SNINT}).
The integration over $r$ can be easily carried out and yields a 
confluent hypergeometric function,
\begin{equation}
S_N(\xi;s)=\frac{2^{N-1}(N-1)!}{\pi^{2N}|s|^N}\,
\Phi\left(N, \frac{1}{2}; -\frac{\xi^2}{|s|}\right).
\label{SN}
\end{equation}
The  function $S_N$ depends on $\alpha_j$, 
$X$, $\bbox{\theta}$, and $\bbox{\psi}$ only through a specific 
combination $\xi=X-\tilde{X}$.  

The parameter $s$ must be negative because the integral (\ref{SNINT})
would diverge otherwise. This implies that only smoothed Wigner 
functions corresponding to $s<0$ can be directly sampled 
from homodyne statistics. The confluent hypergeometric functions can be 
expressed in terms of the error function of the imaginary argument 
${\rm erfi}(x)$. It holds that

\end{multicols}
\begin{figure}
\centerline{\epsfig{figure=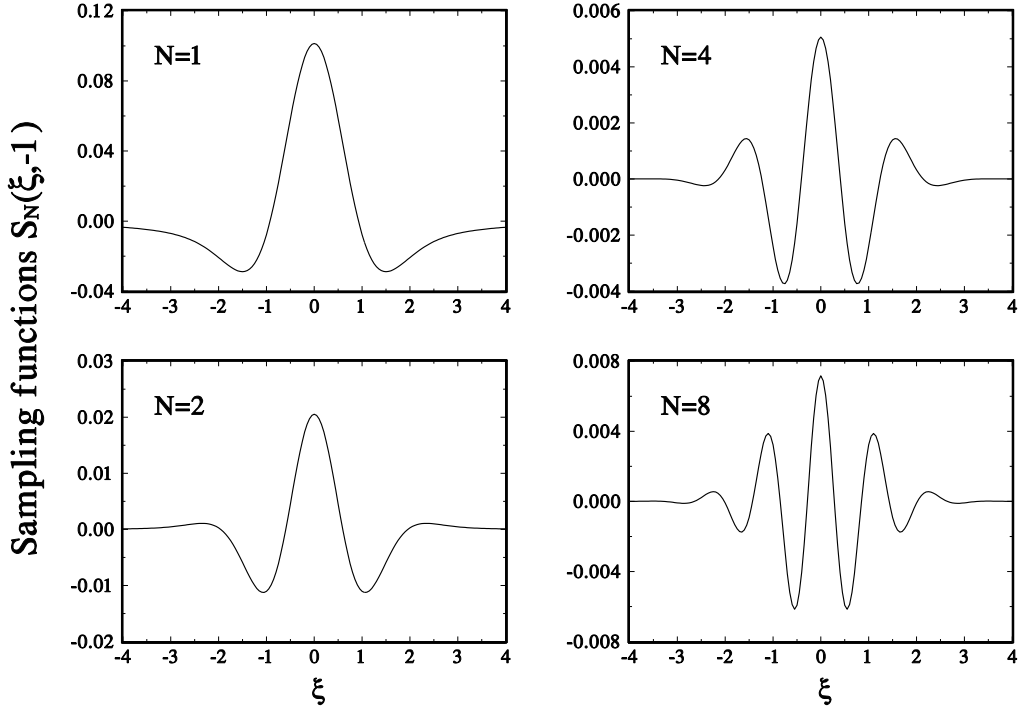,width=0.75\linewidth}}
\vspace*{3mm}
\caption{Sampling functions $S_N(\xi,-1)$
 for the Husimi $Q$-function of $N$-mode optical field. }
\end{figure}
\vspace*{2mm}
\begin{multicols}{2}
\noindent
\begin{eqnarray*}
\Phi\left(1, \frac{1}{2}; -x^2 \right)&=&1-\sqrt{\pi}x e^{-x^2}{\rm erfi}(x),
\\
\Phi\left(N+1, \frac{1}{2}; -x^2 \right)&=&
\frac{(-1)^{N}}{2^{2N} N!}\frac{d^{2N}}{dx^{2N}}
\Phi\left(1, \frac{1}{2}; -x^2 \right),
\end{eqnarray*}
which allows for an easy determination of the required sampling function.

Our results form a multimode generalization of the 
single-mode relations obtained by Vogel and Risken \cite{Vogel89} and 
by Richter \cite{Richter99}. 
Notice also, that D'Ariano {\em et al.} 
gave explicit formula for sampling function 
of two-mode Husimi quasidistribution \cite{DAriano99}.
Several  functions $S_N(\xi;-1)$ are plotted in Fig. 3.
The number of oscillations of $S_N(\xi;s)$ increases with increasing $N$
and the sampling functions are bounded, $S_N\rightarrow 0$  
as $|\xi|\rightarrow \infty$.

\subsection{Imperfect detection and loss-compensating sampling functions}

The sampling functions (\ref{SN}) would yield correct results only 
in the ideal case of unit detection efficiency.
In a realistic experiment, losses are inevitable and the overall detection
efficiency $\eta$ is lower than $1$. The losses can be modeled as a mixing 
of the signal mode with a vacuum on a beam splitter.
The detected quadrature $X^\prime$ is thus a superposition of 
the original quadrature $X$ and a vacuum-state quadrature $X_{\rm vac}$
\cite{Vogel93},
\begin{equation}
X^\prime=\sqrt{\eta}X+\sqrt{1-\eta}X_{\rm vac}.
\label{XPRIME}
\end{equation}
With the help of (\ref{XPRIME}) one can find a simple relation between 
the characteristic functions of the quadratures $X$ and $X^\prime$,
\begin{equation}
\left\langle 
\exp\left(i\sqrt{2}rX\right)\right\rangle=
\exp\left(\frac{1-\eta}{2\eta}r^2\right)
\left\langle 
\exp\left(i\sqrt{\frac{2}{\eta}} r X^{\prime}\right)\right\rangle.
\label{XANDXPRIME}
\end{equation}
Inserting formula (\ref{XANDXPRIME}) into (\ref{CS}) and repeating the steps 
leading to Eq. (\ref{SN}) one finds that the replacements
\begin{equation}
s\rightarrow s+\frac{1-\eta}{\eta},
\qquad
X\rightarrow \frac{X}{\sqrt{\eta}}
\label{REPLACEMENT}
\end{equation}
are necessary and sufficient in Eq. (\ref{WSSAMPLED}) to account 
for detection losses,
\begin{equation}
S_N\left(X,\tilde{X};s,\eta\right)=
S_N\left(\frac{X}{\sqrt{\eta}}-
\tilde{X};s+\frac{1-\eta}{\eta}\right).
\label{SNLOSS}
\end{equation}
The losses impose a new limit on the ordering parameter $s$
because the modified ordering parameter $s+(1-\eta)/\eta$ must be negative,
\begin{equation}
s<-\frac{1-\eta}{\eta}\equiv s_\eta.
\label{SBOUND}
\end{equation}
Only smoothed Wigner functions with $s<s_\eta$ can be reconstructed
if losses are present in the experiment.

\section{Density matrix elements}

In this section we briefly address the sampling of  multimode 
density matrix elements in the Fock state basis
$|\{n_l\}\rangle=|n_1\rangle|n_2\rangle\ldots|n_N\rangle$,
\begin{equation}
\rho_{{\bf m}{\bf n}}=\langle \{m_l\}|\rho
|\{n_l\}\rangle,
\end{equation}
where ${\bf m}=m_1,\ldots,m_N$ and ${\bf n}=n_1,\ldots,n_N$
are vector indices used for notation simplicity.
In tomography with single LO the matrix elements  $\rho_{\bf mn}$ 
can be reconstructed from the measured data according to
\begin{eqnarray}
&&\rho_{{\bf m}{\bf n}}=
\int_{\Omega} \,d\Omega \int_{-\infty}^{\infty}\,dX 
f_{\bf m n}(X,\bbox{\theta},\bbox{\psi})
w(X;\bbox{\theta},\bbox{\psi}).
\label{RHO}
\end{eqnarray}
The functions $f_{\bf m n}$ 
were expressed in terms of integrals in Ref.  \cite{DAriano99}. 
Well-known analytical formulas for single-mode sampling 
functions $f_{mn}$ involve products of regular and irregular eigenfunctions 
of the harmonic oscillator Hamiltonian \cite{Leonhardt96,Richter00}. The 
two-mode functions $f_{m_1 m_2,n_1 n_2}$ can be written 
as finite series of the confluent hypergeometric functions \cite{Richter97}.
Here we show how to derive analytical expressions for arbitrary
sampling functions $f_{{\bf m n}}$ for generic $N$-mode optical field. 
Our starting point shall be  multimode Husimi $Q$-function,
\begin{equation}
Q(\{\alpha_j\})=\frac{1}{\pi^N}\langle\{\alpha_j\}|\rho|\{\alpha_j\} \rangle,
\label{Q}
\end{equation}
where $|\{\alpha_j\}\rangle$ is multimode coherent state. 
When the density operator $\rho$ is expanded in Fock  basis
the Eq. (\ref{Q}) takes the form
\end{multicols}
\vspace*{-0.5\baselineskip}
\noindent\rule{0.5\textwidth}{0.4pt}\rule{0.4pt}{0.5\baselineskip}
\vspace*{3mm}
\begin{eqnarray}
Q(\{\alpha_j\})=\frac{1}{\pi^N}
\sum_{m_1,n_1=0}^{\infty}\ldots\sum_{m_N,n_N=0}^{\infty}
\rho_{{\bf m}{\bf n}}
\prod_{j=1}^N
\frac{\alpha_j^{\ast m_j}\alpha_j^{n_j}}{\sqrt{m_j!\,n_j!}} e^{-|\alpha_j|^2}.
\end{eqnarray}
From this expansion we can readily see that Husimi quasidistribution
$Q(\{\alpha_j\})\equiv W_{-1}(\{\alpha_j\})$ is a generating
function of the density matrix elements in Fock basis,
\begin{eqnarray}
\rho_{\bf m n}=
\pi^N \prod_{j=1}^N
\frac{1}{\sqrt{m_j!\, n_j!}}
\frac{\partial^{m_j}}{\partial \alpha_j^{\ast m_j}}
\frac{\partial^{n_j}}{\partial\alpha_{j}^{n_j}}
\left.\left[Q(\{\alpha_j\})\prod_{l=1}^N e^{|\alpha_l|^2}\right]
\right|_{\alpha_j=\alpha_j^{\ast}=0}.
\label{RHOQ}
\end{eqnarray} 
It follows immediately that the sampling function for the Husimi 
quasidistribution is a generating function of the sampling functions 
$f_{\bf m n}$. This can be shown explicitly 
by inserting the expressions (\ref{WSSAMPLED}) and (\ref{RHO}) 
into Eq. (\ref{RHOQ}) and comparing left- and right-hand sides 
of the resulting formula. We have
\begin{equation}
f_{{\bf m n}}(X,\bbox{\theta},\bbox{\psi};\eta)=
\pi^N \prod_{j=1}^N
\frac{1}{\sqrt{m_j!\, n_j!}}
\frac{\partial^{m_j}}{\partial \alpha_j^{\ast m_j}}
\frac{\partial^{n_j}}{\partial\alpha_{j}^{n_j}}
\label{FGEN} 
\left.\left[
S_N\left(X,
\tilde{X}(\{\alpha_j\},\bbox{\theta},\bbox{\psi});s=-1,\eta\right)
\prod_{l=1}^N e^{|\alpha_l|^2}\right]
\right|_{\alpha_j=\alpha_j^{\ast}=0}.
\label{FOCKGENER}
\end{equation} 
\vspace*{3mm}

\hspace*{\fill}\rule[0.4pt]{0.4pt}{0.5\baselineskip}%
\rule[0.5\baselineskip]{0.5\textwidth}{0.4pt}
\vspace*{-0.5\baselineskip}
\begin{multicols}{2}
This expression  is general, i.e. valid for any number 
of modes. 
The $Q$-function can be sampled only if the 
detection efficiency $\eta>0.5$, c.f. Eq. (\ref{SBOUND}).
This also limits  the possibility of sampling the density matrix
elements; the functions $f_{{\bf m n}}$
 exist only for $\eta>0.5$.

The dependence of $f_{\bf mn}$ on phases $\psi_j$
can be seen from Eq. (\ref{FGEN}) even without going into explicit calculations.
With the help of the substitution $\alpha_j=\gamma_j\exp(i\psi_j)$ 
one obtains
\begin{equation}
f_{{\bf m n}}(X,\bbox{\theta},\bbox{\psi};\eta)=
F_{{\bf m n}}(X,\bbox{\theta};\eta)
\prod_{j=1}^N e^{i(m_j-n_j)\psi_j},
\end{equation}
moreover, $F_{\bf m n}(X,\bbox{\theta};\eta)$
are real functions. Analytical formula for these so-called pattern functions
$F_{\bf mn}$ can be derived if one inserts the sampling function $S_N$
(\ref{SN})  into (\ref{FOCKGENER})
and performs the necessary differentiations. After a tedious
but straightforward calculation one finds that $F_{\bf mn}$ can be written 
in terms of finite series of confluent hypergeometric functions,
\begin{eqnarray}
F_{\bf mn}(X,\bbox{\theta};\eta)&=&
\frac{2^{N-1}}{\pi^N} \left(\frac{\eta}{2\eta-1}\right)^N
\nonumber \\ && \times
\prod_{j=1}^N \sqrt{\frac{\nu_j!}{\mu_j!}} 
\left[\sqrt{\frac{2\eta}{(2\eta-1)}}u_j(\bbox{\theta})\right]^{\mu_j-\nu_j}
\nonumber \\
&&\times\sum_{k_1=0}^{\nu_1}\ldots \sum_{k_N=1}^{\nu_N}
\Xi_N\left(\frac{X}{\sqrt{2\eta-1}},p_{\bbox{\mu},\bbox{\nu},{\bf k}}\right)
\nonumber \\ &&\times
\prod_{l=1}^N 
\frac{1}{k_l!}{ \mu_l \choose\nu_l-k_l}
\left[\frac{2\eta}{(2\eta-1)} u_l^2(\bbox{\theta})\right]^{k_l},
\nonumber \\
\label{FEXPL}
\end{eqnarray}
where  $\mu_j={\rm max}(m_j,\,n_j)$, $\nu_j={\rm min}(m_j,\,n_j)$,
\begin{equation}
p_{\bbox{\mu},\bbox{\nu},{\bf k}}=\sum_{j=1}^N \mu_j-\nu_j+2k_j,
\end{equation}
and
\begin{eqnarray*}
\Xi_N(x,2k)&=&(-1)^k (N+k-1)! \, \Phi\left(N+k,\frac{1}{2};-x^2\right),
\nonumber \\
\Xi_N(x,2k+1)&=&   2x (-1)^{k} (N+k)! \,
             \Phi\left(N+k+1,\frac{3}{2};-x^2\right).
\end{eqnarray*}
Notice an interesting analogy. The quantum state 
is uniquely and completely determined by its Husimi quasidistribution, 
which contains complete information on all density matrix 
elements $\rho_{\bf mn}$.  Similarly, all the sampling functions 
for density matrix elements can be obtained from the sampling function $S_N$,
which contains all information on $f_{{\bf m n}}$. 

A single-mode version of the formula (\ref{RHOQ}) was
used in  \cite{DAriano94} to find the sampling functions for 
single-mode density matrix elements. However, 
the sampling function $S_N$ was not explicitly given in \cite{DAriano94} and 
the results were written in form of complicated series. 
Thus later different techniques have been adopted to calculate 
 $F_{mn}$ \cite{Leonhardt95,Richter96b,Leonhardt96,Richter00}.
We emphasize that for $N=1$ the Eq. (\ref{FEXPL}) yields exactly the 
single-mode pattern functions  $F_{mn}$ given in \cite{Leonhardt95,Leonhardt96} 
and for $N=2$ we get the two-mode pattern functions derived in 
\cite{Richter97}.
The formula (\ref{FEXPL}) is also suitable for  investigation 
of the asymptotic  behavior.  One  simply inserts
the asymptotic expansions of relevant confluent hypergeometric functions 
into (\ref{FEXPL}) and extracts the asymptotic expansion of
$F_{{\bf m n}}$. It turns out that all pattern functions are bounded
and go to zero as $|X|\rightarrow\infty$.

Finally  we note that the functions $F_{{\bf m n}}$ 
given by (\ref{FEXPL})
 differ from those obtained in Ref. \cite{DAriano99}
because  we have removed  a superfluous imaginary part of 
 $S_N$. Had we retained this imaginary part, we would have
obtained the pattern functions derived in \cite{DAriano99}. To see this, 
one can insert the integral representation (\ref{SNINT}) into Eq. (\ref{FGEN})
and differentiate prior to the integration. One recovers the formula
(13) of  Ref. \cite{DAriano99},
\begin{eqnarray}
F_{\bf mn}(X,\bbox{\theta};\eta)&=&\frac{1}{\pi^N}\prod_{j=1}^N 
\sqrt{\frac{\nu_j!}{\mu_j!}}\,[-i u_j(\bbox{\theta})]^{\mu_j-\nu_j}
\nonumber \\
&&\times\int_0^\infty  dr \, e^{\frac{1-2\eta}{2\eta}r^2} 
e^{i\sqrt{2/\eta}rX}r^{2N-1}
\nonumber \\
&&\times\prod_{l=1}^N  r^{\mu_l-\nu_l}
L_{\nu_l}^{\mu_l-\nu_l}[r^2 u_l^2(\bbox{\theta})],
\label{FRHOPAT}
\end{eqnarray}
where $L_n^\alpha(x)$ denotes generalized Laguerre polynomial.
A real part of the complex function (\ref{FRHOPAT})
coincides with Eq. (\ref{FEXPL}).

\begin{figure}
\centerline{\epsfig{figure=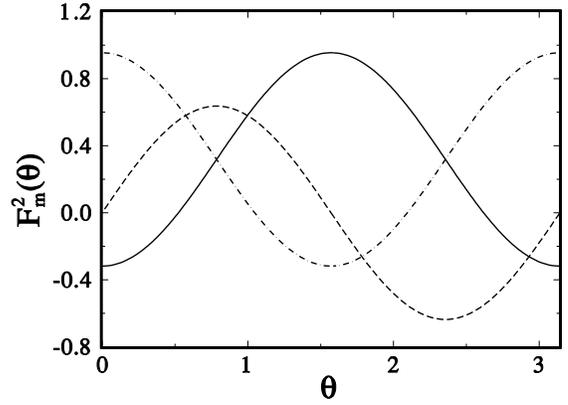,width=0.85\linewidth}}
\vspace*{3mm}
\caption{The functions $F_m^2(\theta)$ biorthogonal to
$G_k^2(\theta)$ in the interval $[0,\pi]$;
 $m=0$ solid line, $m=1$ dashed line, $m=2$ dot-dashed line. }
\end{figure}

\section{$S$-ordered moments}

\subsection{Multimode sampling functions}

Here we consider direct sampling of the multimode $s$-ordered moments
\begin{equation}
C_{\bf m n}^{(s)}=
\langle a_1^{\dagger m_1}\ldots a_N^{\dagger m_N}
 a_1^{n_1}\ldots a_{N}^{n_N}\rangle_s .
\label{C}
\end{equation}
We will follow an approach due to Opatrn\'{y} {\em et al.} \cite{Opatrny97a}
and generalize their results 
for two-mode homodyning to  any number of modes. The quadrature distribution
$w(X;\bbox{\theta},\bbox{\psi})$ can be obtained from the joint distribution
$w(x_1,\ldots,x_N;\bbox{\psi})$ of the single mode quadratures 
\begin{equation}
x_j=\frac{1}{\sqrt{2}}(a_je^{-i\psi_j}+a_j^{\dagger}e^{i\psi_j})
\end{equation}
according to
\begin{eqnarray}
w(X;\bbox{\theta},\bbox{\psi})&=&
\int_{-\infty}^\infty \, d x_1\, \ldots \int_{-\infty}^\infty dx_N \,
w(x_1,\ldots,x_N; \bbox{\psi})
\nonumber \\
&&\times\delta\left(X-\sum_{j=1}^N u_j(\bbox{\theta})x_j\right).
\label{WXDELTA}
\end{eqnarray}
The moments (\ref{C}) can be reconstructed from the joint quadrature statistics 
as follows:
\begin{eqnarray}
C_{\bf m n}^{(s)} &=&
\int w(x_1,\ldots,x_N;\bbox{\psi})
\prod_{j=1}^N  \, d x_j \, d\psi_j \,
\left(\frac{s}{2}\right)^{(m_j+n_j)/2}
\nonumber \\
&&\times
H_{m_j+n_j}\left(\frac{x_j}{\sqrt{s}}\right)
K(m_j,n_j) e^{i(n_j-m_j)\psi_j},
\label{CSMJOINT}
\end{eqnarray}
where we integrate over $N$ quadratures $x_j\in(-\infty,\infty)$
and $N$ phases $\psi_j\in [0,\pi]$.
$H_n(x)$ denotes customary Hermite polynomial of variable $x$ and
\begin{equation}
K(m,n)=\left[ \pi {m+n \choose n}\right]^{-1}.
\end{equation}
The multimode sampling function employed in Eq. (\ref{CSMJOINT}) 
is just a product of the appropriate single-mode sampling functions 
derived in \cite{Richter99a}.

We would like to link $C_{\bf m n}^{(s)}$ 
to the quadrature distribution 
$w(X;\bbox{\theta},\bbox{\psi})$, 
\begin{equation}
C_{\bf m  n}^{(s)}=
\int_{\tilde{\Omega}} \, d\tilde{\Omega} \int_{-\infty}^{\infty} \, d X
D_{\bf m n}(X;\bbox{\theta},\bbox{\psi};s)
w(X;\bbox{\theta},\bbox{\psi}),
\label{CSLO}
\end{equation}
where
\begin{equation}
\int_{\tilde{\Omega}} d\tilde{\Omega}=
\prod_{l=1}^{N-1}
\int_{0}^{\theta_{\rm max}} d\theta_l
\prod_{j=1}^{N} \int_{0}^{\pi} d\psi_j.
\end{equation}
Notice the definition interval of the phase variables, $\psi_j\in[0,\pi]$. 
The upper bound of integration over $\theta_j$ 
is denoted by $\theta_{\rm max}$.
The most straightforward choice would be, of course, to keep
$\theta_{\max }=\pi/2$ as in previous sections.
As we shall see later, the choice $\theta_{\rm max}=\pi$ can be more suitable. 

Following \cite{Opatrny97a} we shall look for the sampling function 
$D_{\bf m n}$ in the factorized form,
\begin{eqnarray}
D_{\bf m n}(X,\bbox{\theta},\bbox{\psi};s)&=&\left(\frac{s}{2}\right)^{M/2}
H_M\left(\frac{X}{\sqrt{s}}\right)
\prod_{l=1}^{N-1} F_{m_l+n_l}^{M_l}(\theta_l)
\nonumber \\
&&\times \prod_{j=1}^N K(m_j,n_j)e^{i(n_j-m_j)\psi_j},
\label{DIDEAL}
\end{eqnarray}
where $M=\sum_{l=1}^N(m_l+n_l)$ and  $F_{m_l+n_l}^{M_l}(\theta_l)$ 
are some yet undetermined functions.
Inserting Eqs. (\ref{DIDEAL}) and (\ref{WXDELTA}) into Eq. (\ref{CSLO})  and 
comparing the resulting expression with (\ref{CSMJOINT}) we conclude 
that the following integral equation must be fulfilled:
\end{multicols}
\vspace*{-0.5\baselineskip}
\noindent\rule{0.5\textwidth}{0.4pt}\rule{0.4pt}{0.5\baselineskip}
\vspace*{3mm}
\begin{eqnarray}
\int_{0}^{\theta_{\rm max}}\,d\theta_1\ldots\int_0^{\theta_{\rm max}}
\, d\theta_{N-1}\,
H_M\left(\frac{1}{\sqrt{s}}\sum_{j=1}^N x_j u_j(\bbox{\theta})\right)
\prod_{l=1}^{N-1} F_{m_l+n_l}^{M_l}(\theta_l)
=\prod_{j=1}^N H_{m_j+n_j}\left(\frac{x_j}{\sqrt{s}}\right).
\label{INTEQ}
\end{eqnarray}
\vspace*{3mm}

\hspace*{\fill}\rule[0.4pt]{0.4pt}{0.5\baselineskip}%
\rule[0.5\baselineskip]{0.5\textwidth}{0.4pt}
\vspace*{-0.5\baselineskip}
\begin{multicols}{2}
We shall need the  summation rule for Hermite polynomials,
\begin{equation}
H_{l}(x_1\cos\theta+x_2\sin\theta)=\sum_{k=0}^l
G_{k}^l(\theta)H_k(x_1)H_{l-k}(x_2),
\label{HSUM}
\end{equation}
where
\begin{equation}
G_{k}^l(\theta)={l \choose k}(\cos\theta)^k (\sin\theta)^{l-k}.
\label{G}
\end{equation}
If we use the summation rule (\ref{HSUM}) repeatedly we find that
\begin{eqnarray}
&&H_M\left(\frac{1}{\sqrt{s}}\sum_{j=1}^N x_j u_j(\bbox{\theta})\right)
=
\nonumber \\
&&\qquad{\sum_{j_1,\ldots,j_N}}^{\hspace*{-2mm}\prime} \;
H_{j_N}\left(\frac{x_N}{\sqrt{s}}\right)
\prod_{l=1}^{N-1} G_{j_l}^{k_l}(\theta_l)H_{j_l}
\left(\frac{x_l}{\sqrt{s}}\right),
\label{HEXPAND}
\end{eqnarray}
which is a multimode generalization of (\ref{HSUM}).
\noindent
The prime denotes sum over all $j_1,\ldots,j_N$
meeting the constraint $\sum_{l=1}^N j_l=M$, and
\begin{equation}
k_l=\sum_{p=l}^N j_p .
\label{KL}
\end{equation}
The  expansion (\ref{HEXPAND}) is inserted  into  Eq. (\ref{INTEQ})
where the integration over $\theta_l$ should select
the right sequence of the Hermite polynomials.
Let us assume that the functions
 $F_{m}^l(\theta)$  are
biorthogonal to $G_{k}^l(\theta)$ in the interval $[0,\theta_{\rm max}]$,
\begin{equation}
\int_{0}^{\theta_{\rm max}} \, d\theta \,
G_{k}^l(\theta)F_{m}^l(\theta) =\delta_{m,k},\qquad
k=0,\ldots,l.
\label{FGORT}
\end{equation}
 The biorthogonality property (\ref{FGORT}) ensures
that the integral equation (\ref{INTEQ}) is fulfilled
if the indices $M_l$ are constructed in the same way as $k_l$, Eq. (\ref{KL}),
where $j_p$ is replaced by $m_p+n_p$,
\begin{equation}
M_l=\sum_{p=l}^N m_p+n_p.
\end{equation}
Indeed,  the integration over $\theta_1$ in (\ref{INTEQ}) then
selects correct value of the sum $m_1+n_1$, subsequent integration over 
$\theta_2$ fixes $m_2+n_2$ and so on.
Notice also that the correct values of the differences $n_j-m_j$ are fixed
by the exponentials $\exp[i(n_j-m_j)\psi_j]$ in (\ref{DIDEAL}).
The sampling functions for multimode $s$-ordered moments are thus given
by formula (\ref{DIDEAL}).

The functions $F_m^l(\theta)$ biorthogonal to $G_k^l(\theta)$
have been discussed in \cite{Opatrny97a}.
One can construct them e.g. as linear combinations
of $G_k^l(\theta)$,
\begin{equation}
F_m^l(\theta)=\sum_{k=0}^l A_{mk}^l G_{k}^l(\theta).
\label{FGEXP}
\end{equation}
From the orthogonality conditions (\ref{FGORT}) one obtains a system of
linear equations for the coefficients $A_{mk}^l$, which can be solved 
for each $m$ and $l$.
If we  choose $\theta_{\rm max}=\pi$,
we can find simple analytical formulas for the functions $F_m^l$,
 \begin{equation}
 F_m^l(\theta)=\sum_{k=0}^l e^{i(l-2k)\theta} E_{mk}^l,
 \label{FORT}
 \end{equation}
 where (see Appendix for derivation),
 \begin{equation}
 E_{mk}^l=\frac{i^{l-m}}{\pi}{l \choose k }^{-1}\sum_{j=0}^{m}
 {m \choose j} {l-m \choose k-j} (-1)^{k-j}.
 \label{EFORT}
 \end{equation}
The functions $F_m^2(\theta)$ are plotted in Fig. 4.

The sampling functions compensating imperfect detection
 $\eta<1$, can be obtained from (\ref{DIDEAL})
by making use of the simple replacement (\ref{REPLACEMENT}),
\begin{equation}
D_{\bf m n}(X,\bbox{\theta},\bbox{\psi};s,\eta)=
D_{\bf m n}
\left(\frac{X}{\sqrt{\eta}},\bbox{\theta},\bbox{\psi};
s+\frac{1-\eta}{\eta}\right).
\label{DLOSS}
\end{equation}
This relation is particularly simple when normally ordered moments 
are considered \cite{Richter96}. Inserting $s=1$ into (\ref{DLOSS}) we have
\begin{equation}
D_{\bf m n}(X,\bbox{\theta},\bbox{\psi};1,\eta)=
\eta^{-M/2} D_{\bf m n}\left(X,\bbox{\theta},\bbox{\psi};1\right).
\label{DLOSSNORMAL}
\end{equation}
Normally ordered moments do not contain any contribution 
from vacuum fluctuations and they all vanish for a vacuum state. 
The experimental losses effectively reduce the value of normally ordered moment
of $M$-th order by a factor $\eta^{M/2}$. To compensate for 
imperfect detection, it suffices to use the ideal sampling function
as if the detection was perfect and then divide the result
by $\eta^{M/2}$.

\subsection{Effect of aliasing and reconstruction limits}

In the experiment, the statistics $w(X;\bbox{\theta},\bbox{\psi})$
 are measured only at a certain finite number 
of angles $\theta_j^{(k)}$ and phases $\psi_j^{(k)}$ and
the integration  over $d\tilde{\Omega}$ 
is replaced by a summation over  finite number of
discrete points $(\bbox{\theta},\bbox{\psi})$. 
This discretization imposes  limits
on the order  of the reconstructed moments \cite{Wunche96}.

Let us first consider the phases $\psi_j$. To simplify the discussion 
as much as possible, we restrict ourselves for a while to the single-mode case
and sampling of symmetrically ordered moments ($s=0$, Weyl ordering). 
Let us further assume that  the exact quadrature statistics are known for each 
of $N_\psi$ phases $\psi^{(k)}=k\pi/N_\psi$.
The sampling then reads, 
\begin{eqnarray}
&&\langle a^{\dagger m}a^n\rangle_{\rm sym}
=\frac{\pi}{N_{\psi}} {2}^{(m+n)/2}K(m,n)
\nonumber \\
&&\quad\times\sum_{k=1}^{N_\psi} \exp\left(i(n-m)\frac{k\pi}{N_\psi}\right)
\int_{-\infty}^\infty d x \, x^{m+n} w\left(x, \frac{k\pi}{N_{\psi}}\right).
\nonumber \\
\label{SMRECONSTRUCT}
\end{eqnarray}
The formula (\ref{SMRECONSTRUCT}) is  a discrete Fourier transform 
in $\psi$. The Fourier series of the quadrature moments,
\begin{eqnarray}
&&\int_{-\infty}^\infty d x \, x^{m+n} w\left(x,\psi\right)
= 2^{-(m+n)/2}
\nonumber \\
&&\quad\times\sum_{k=0}^{m+n}{m+n \choose k}
\left\langle a^{\dagger m+n-k}a^{k}\right\rangle_{\rm sym}
e^{i(m+n-2k)\psi},
\label{XMOMEXP}
\end{eqnarray}
contains either  odd or  even frequencies
depending on the parity of $m+n$.
The $N_\psi$-point discrete Fourier transform (\ref{SMRECONSTRUCT})
gives correct results only for sufficiently low moments, because it 
cannot discriminate between $\exp[ik\psi]$ and 
$\exp[i(k+2N_{\psi})\psi]$. This phenomenon is called aliasing 
\cite{Leonhardt97b} and  it imposes an upper bound on the order of the
reconstructed moment. When we  substitute Fourier expansion (\ref{XMOMEXP}) 
into Eq. (\ref{SMRECONSTRUCT}), we find that $m<N_{\psi}$ and $n<N_\psi$
must hold simultaneously.
The same limitation obviously holds for any $s$-ordering and 
also for multimode moment reconstruction with sampling functions $D_{\bf mn}$.
In particular,
\begin{equation}
m_j<N_{\psi_j},\qquad n_j<N_{\psi_j},
\label{MNVARPHI}
\end{equation}
must be fulfilled, where $N_{\psi_j}$ is the number of sampling points 
of the phase $\psi_j$.

Let us proceed to the angles $\theta_j$.
For a successful reconstruction, it is crucial to meet  the biorthogonality 
conditions (\ref{FGORT}) where the integration is replaced by 
summation over $N_\theta$ angles $\theta^{(n)}$,
\begin{equation}
\sum_{n=1}^{N_{\theta}} F_{m}^l(\theta^{(n)}) G_k^l(\theta^{(n)})=\delta_{mk},
\quad m,k=0,\ldots,l.
\label{FDISCRETE}
\end{equation}
If the condition (\ref{FDISCRETE}) is violated due to discretization,
then the reconstruction could be spoiled with large systematic error 
and the sampling would not yield reliable results. 
We shall prove below that the functions (\ref{FORT})
fulfill the conditions (\ref{FDISCRETE}) provided that the sampling points 
are equidistant, $\theta^{(n)}=n\pi/N_{\theta}$, and $l<N_\theta$ holds.

First of all we recall that the functions $G_k^l(\theta)$, Eq. (\ref{G}),
and $F_m^l(\theta)$, Eq. (\ref{FORT}), can be expanded in finite 
Fourier series,
with the highest component equal to $l$ in both cases. Moreover, both
functions contain only odd or only even Fourier components,
depending on the parity of $l$. 
If  $F_m^l(\theta)$ is expanded in Fourier series,
then Eq. (\ref{FDISCRETE}) becomes a summation of several discrete 
Fourier transforms of $G_k^l(\theta)$ (we assume
$\theta^{(k)}=k\pi/N_{\theta}$).
If $l<N_\theta$, then all discrete Fourier transforms yield the
same results as the original integrations, and (\ref{FDISCRETE}) 
holds exactly. 
The main advantage of the choice $\theta_{\rm max}=\pi$ is now clear.
It has allowed us to find analytical expressions for the functions
$F_m^l$ which meet the discretized biorthogonality conditions 
(\ref{FDISCRETE}). 
We remark that the functions $F_{m,n}^l\equiv F_m^l(\theta^{(n)})$ 
can also be constructed numerically by solving a system of Eqs. 
(\ref{FDISCRETE})  for a given set of sampling points $\theta^{(n)}$
\cite{McAlister97}. Looking at formula (\ref{DIDEAL}) we find 
the limit on the order of reconstructed multimode moments,
 \begin{equation}
 M_j<N_{\theta_j}.
 \label{MNTHETA}
 \end{equation}

We can conclude that if the quadrature statistics 
$w(X;\bbox{\theta},\bbox{\psi})$ are  measured with high accuracy,
then sampling at finite number of points $(\bbox{\theta},\bbox{\psi})$  
provides sufficient information for the successful reconstruction 
of certain $s$-ordered moments $C_{\bf mn}^{(s)}$.
If we use the sampling functions $D_{\bf mn}$ and we want to reconstruct
all moments of $M$th order we have to measure $w(X;\bbox{\theta},\bbox{\psi})$
at 
\begin{equation}
R(M,N)=(M+1)^{2N-1}
\end{equation}
 points $(\bbox{\theta},\bbox{\psi})$ 
(we have the factor $M+1$ for each of $N$ phases $\psi_j$ and 
$N-1$ angles $\theta_l$). 
This number of sampling points  is sufficient, but not necessary.
The  $M$th order moments of $N$-mode field 
can be parametrized by $P(M,N)$ real numbers, where
\begin{equation}
P(M,N)={ M+2N-1\choose M }.
\end{equation}
It suffices to measure the statistics
$w(X;\bbox{\theta},\bbox{\psi})$ at $P(M,N)$ distinct points 
$(\bbox{\theta},\bbox{\psi})$. The appropriate sampling functions must 
be constructed numerically for a given set of sampling points 
$(\bbox{\theta},\bbox{\psi})$ by inverting a system of linear equations
which relates the moments $C_{\bf mn}^{(s)}$ to the moments of the
quadrature statistics $w(X;\bbox{\theta},\bbox{\psi})$.
This approach requires less sampling points because $P(M,N)<R(M,N)$.
Though many interesting questions are related to 
this method, e.g. how to choose the points $(\bbox{\theta},\bbox{\psi})$,
we do not deal with it in this paper in any more detail.

\begin{figure}
\vspace*{10mm}

\centerline{\epsfig{figure=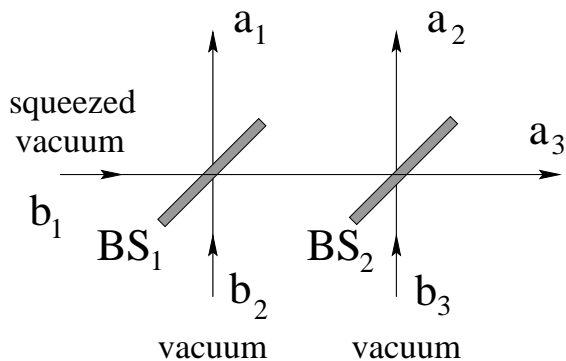,width=0.85\linewidth}}
\vspace*{3mm}
\caption{Preparation of three-mode state. The squeezed vacuum in mode $b_1$
is mixed on beam splitters BS$_1$ and BS$_2$ 
with two vacua (modes $b_2$ and $b_3$),
yielding the modes $a_1$, $a_2$, and $a_3$ at the outputs.}
\vspace*{7mm}
\end{figure}

Finally, we should note that the sampling functions are not 
unique. This is a general feature of optical homodyne tomography. 
An infinite number of
functions exist whose average over $w(x;\bbox{\theta},\bbox{\psi})$ is zero for all
physical quadrature distributions. These so-called null functions
can be freely added to the above derived sampling functions. 
This freedom of choice is exploited in
adaptive homodyne tomography to find the sampling functions 
minimizing statistical error for a given set of experimental data
\cite{DAriano99b}.

\begin{figure}
\vspace*{3mm}
\centerline{\epsfig{figure=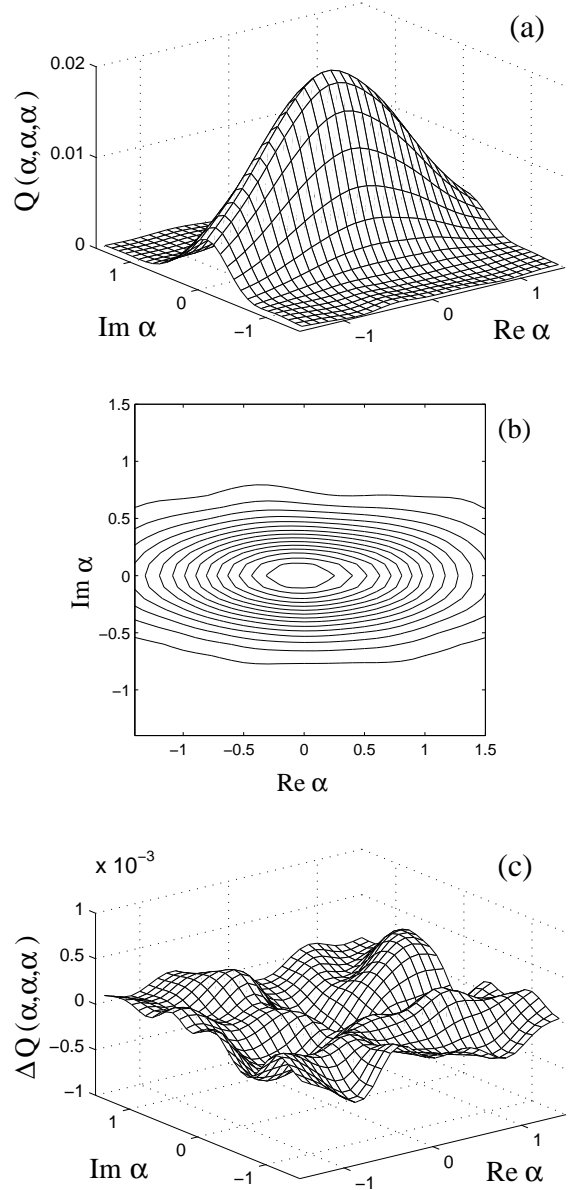,width=0.85\linewidth}}
\vspace*{3.5mm}
\caption{Reconstruction of the three-mode $Q$-function of a squeezed state
prepared according to Fig. 5. 
A two-dimensional cut $Q(\alpha,\alpha,\alpha)$
through the six-dimensional phase space is plotted. Shown are surface
(a) and contour (b) plots of the reconstructed quasidistribution
and a difference $\Delta Q$ between reconstructed and exact
$Q$-functions (c). }
\end{figure}

\section{Monte Carlo simulations}

We have performed  Monte Carlo simulations of multimode homodyne detection with 
a single LO and tested the performance of the sampling fnctions.  
Since the reconstruction
of multimode density matrix elements was already considered to
relatively large extent in Ref. \cite{DAriano99}, we focus here
on the sampling of the multimode quasidistributions and s-ordered moments.
The main purpose of this section is to illustrate the applicability
of the above derived sampling functions and the feasibility
of successful reconstruction of three-mode quantum state 
from an acceptably large amount of data.

To be more specific, we consider  three-mode squeezed state prepared 
according to Fig. 5. This state represents a simple but nontrivial example
exhibiting nonclassical properties (squeezing). 
As depicted in Fig. 5, single-mode squeezed vacuum in mode $b_1$ 
is mixed on two beam splitters BS$_1$ and BS$_2$
 with two vacua $b_2$ and $b_3$. The output modes
 \begin{eqnarray}
a_1&=&\frac{1}{\sqrt{3}}b_1+\frac{2}{\sqrt{6}} b_2,
\nonumber \\
a_2&=&\frac{1}{\sqrt{3}}b_1-\frac{1}{\sqrt{6}} b_2-\frac{1}{\sqrt{2}}b_3,
\nonumber \\
a_3&=&\frac{1}{\sqrt{3}}b_1-\frac{1}{\sqrt{6}} b_2+\frac{1}{\sqrt{2}}b_3,
\label{ABTRANSFORM}
\end{eqnarray}
 can then enter the multimode homodyne detector shown in Fig. 2.
The transformation (\ref{ABTRANSFORM}) is unitary, thus preserving the
canonical commutation relations. Moreover,
\begin{equation}
b_1=b_0\cosh r+b_0^\dagger \sinh r,
\end{equation}
where $b_0$ is annihilation operator of vacuum state and $r$ is squeezing
parameter. We assume $r=1$ in  the following.

The reconstructed three-mode $Q$-function is shown in Fig. 6. 
In the computer simulation, we have sampled  at $10$   angles 
$\theta_l^{(k)} = k \pi / 20$,  and phases 
$\psi_j^{(k)} = 2 \pi k/10$,  $k = 1,\ldots,10$.
 At each point $(\bbox{\theta},\bbox{\psi})$
  the quadrature has been  measured $50$ times so that
the total amount of acquired data is $5\times 10^6$.
We have assumed a  detection efficiency
$\eta=0.8$ and we have used the loss-compensating sampling function 
(\ref{SNLOSS}).  $Q(\alpha_1,\alpha_2,\alpha_3)$ is a function 
in six-dimensional phase space, it is impossible to plot it as a whole
and we must restrict ourselves to some lower-dimensional subspaces 
of the phase  space. In Figure 6 we show
a two-dimensional cut $Q(\alpha,\alpha,\alpha)$. The reconstructed $Q$-function
exhibits Gaussian shape characteristic for squeezed states. The 
squeezing is clearly reflected in the 
elliptic shape of the $Q$-function, as can be seen in
the contour plot in Fig. 6(b).
The reconstruction error can be judged from Fig. 6(c), which depicts
the difference $\Delta Q$ between exact and reconstructed $Q$-functions.
The error is acceptably small and the reconstruction can be considered 
successful.

Let us now proceed to sampling the multimode moments. 
Again, we  have assumed $\eta=0.8$. We  have employed the loss-compensating 
sampling kernels  (\ref{DLOSSNORMAL}) and the analytical 
functions $F_m^l(\theta)$ given by Eq. (\ref{FORT}). 
We have sampled  at $10$ different 
values of each angle $\theta_l^{(k)}=k\pi/10$ and phase $\psi_j^{(k)}=k\pi/10$,
$k=1,\ldots,10$.  At each point $(\bbox{\theta},\bbox{\psi})$ $200$
values of the quadrature $X$ were recorded, which represents altogether
$2\times 10^7$ data.

The reconstructed  normally ordered moments 
$\langle :\!\!n_1^k\!\!:\rangle$ and $\langle a_1^k\rangle$ can 
be seen in Fig. 7.
The gray bars display the exact values and allow for comparison with 
the sampled moments. The reconstructed moments are in very 
good agreement with the exact values. The sampling error increases 
with the moment order and it is higher for 
 $\langle a_1^k\rangle$ than for the factorial moments
$\langle :\!\!n_1^k\!\!:\rangle\equiv\langle a_1^{\dagger k}a_1^k\rangle$. 
While  $\langle:\!\!n_1^3\!\!:\rangle$
is still reconstructed with high accuracy,  $\langle a_1^6\rangle$
is sampled with certain error. This can be explained by the necessity
of  sampling a high Fourier component $\exp(6i\psi_1)$
in order to reconstruct  $\langle a_1^6\rangle$. 
When we tried to sample moments of $10$th or higher orders,
the results suffered from very large systematic errors because we 
violated the conditions (\ref{MNVARPHI}) and (\ref{MNTHETA}).

The moments $\langle :\!n_1^k\!:\rangle$ contain information on the
pho\-ton-number statistics of the mode $a_1$. 
From the sampled moments we can determine the 
Mandel $Q$-parameter for $j$th mode,
\begin{equation}
Q_j=\frac{\langle :\! (\Delta n_j)^2\!:\rangle}{\langle n_j\rangle}
=\frac{\langle :\! n_j^2\!:\rangle-\langle n_j \rangle^2}{\langle n_j\rangle}.
\end{equation}
This parameter allows one to quickly distinguish between 
super-Poissonian ($Q_j>0$) and sub-Poissonian ($Q_j<0$) photon-number statistics.
From the data shown in Fig. 8 we have $Q_1\approx 1.25$ and we find that
the light in the mode $a_1$ exhibits
super-Poissonian photon number statistics.
 The moments of the modes $a_2$ and $a_3$ 
are the same as those of mode $a_1$ because the squeezed vacuum $b_1$ 
is equally split among the three modes $a_j$, c.f. Eq. (\ref{ABTRANSFORM}).
The sampling works equally well  for the modes $a_2$ and $a_3$ 
and the results are very similar to those displayed in Fig. 7.

\begin{figure}
\centerline{\epsfig{figure=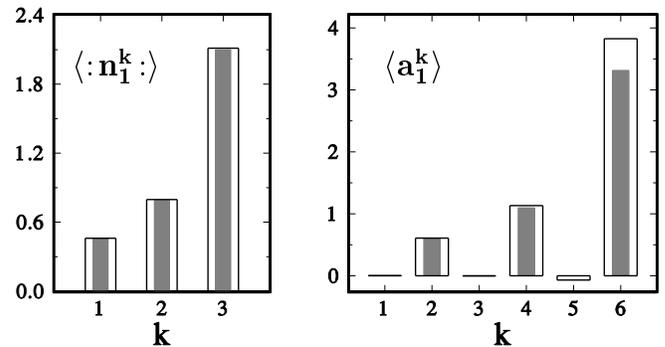,width=0.99\linewidth}}
\vspace*{3mm}
\caption{Sampled moments of the mode $a_1$. The empty solid bars
show the reconstructed moments, the gray bars display exact values 
for comparison. Only real parts of the moments $\langle a_1^k \rangle$
are shown.}
\end{figure}

\begin{figure}
\centerline{\epsfig{figure=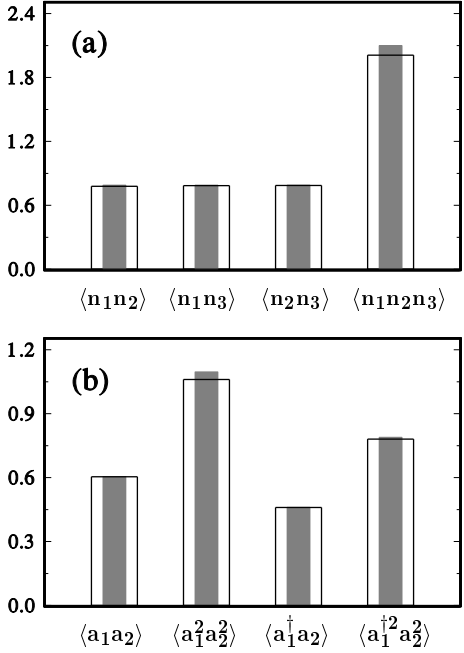,width=0.7\linewidth}}
\vspace*{3mm}
\caption{Sampled two-mode and three-mode moments. In Fig. (b), real
parts of the complex moments are displayed.}
\end{figure}

Having verified the feasibility of reconstruction of the single-mode
moments  we have finally  sampled the multimode moments.
Several results are shown in Fig. 8. Again, the low-order moments
are reproduced with high accuracy, and the error increases with
the moment order. 
It is worth noting that the photon number correlations
\begin{equation}
\langle n_1^{k_1} n_2^{k_2}\ldots n_N^{k_N}\rangle_s
\label{NCORR}
\end{equation}
can be sampled from phase averaged data. This is important 
from the experimental point of view, because the sampling of moments
(\ref{NCORR}) does not require stable relative phase between the local 
oscillator and signal modes. All phases $\psi_j$ can be fully randomized,
e.g., by means of randomly driven piezoelectric modulators,
and the homodyning then yields phase-averaged quadrature statistics 
 \cite{McAlister97}.

In addition to the  squeezed-vacuum state discussed here, 
we have also considered other quantum states, such
as multimode coherent states and multimode
squeezed coherent states. In all cases, the reconstruction procedure
worked well. The numerical simulations clearly demonstrate the 
feasibility of three-mode homodyne tomography 
from  $\approx 10^7$ recorded data. 
Of course, the number of necessary data inevitably increases with the 
number of modes.

\section{Conclusions}

We have derived various important
sampling functions for multimode homodyne tomography with a
single local oscillator.
Starting from the relation between multimode
characteristic function and measured quadrature distribution we have found 
sampling functions for the $s$-parametrized quasidistributions
with $s<s_\eta\leq 0$. We have proved that the sampling function for Husimi
quasidistribution is a generating function of the sampling functions
$f_{\bf mn}$ for density matrix elements in Fock basis $\rho_{\bf mn}$.
The functions $f_{\bf mn}$ were expressed as  finite series of
confluent hypergeometric functions.
Finally, we have found the functions allowing for direct reconstruction
of multimode $s$-ordered moments from the homodyne data.
In all cases, loss-compensating sampling functions, applicable to
a realistic experiment with detection efficiency $\eta<1$, have been provided.
In order to test  performance of the sampling functions we 
simulated homodyne detection of squeezed three-mode state and reconstructed
the three-mode $Q$-function and several normally ordered moments. 
The reconstruction has shown very good results for a  detection efficiency
$\eta=0.8$ and  $10^7$ sampled data, which is experimentally feasible.
We emphasize that the multimode quantum state is reconstructed
from the statistics of a class of single-mode quadratures.
Only one homodyne detector is needed, which substantially simplifies
the experiment. This method is particularly suitable for the
measurement of ultrafast internal correlations of optical pulses
or for the reconstruction of the quantum state of multimode
single-frequency optical field.

\acknowledgments 

The author would like to thank  T. Opatrn\'{y}, J. Pe\v{r}ina, and
D.-G. Welsch for stimulating and helpful discussion. 
Financial support of the U.S.-Israel Binational Science Foundation 
(Grant No. 96-00432) is gratefully acknowledged.
 
\appendix

\section*{}

Here we derive the expression (\ref{FORT}) for the functions $F_m^{l}(\theta)$.
We insert the explicit form (\ref{G}) of the function 
$G(\theta)$ into (\ref{FGORT}), multiply
by $\alpha^k (i\beta)^{l-k}$ and sum over $k$,
\begin{equation}
\sum_{k=0}^l\int_0^{\pi}{l \choose k}(\alpha\cos\theta)^k 
(i\beta\sin\theta)^{l-k} F_m^{l}(\theta) d\theta=\alpha^m(i\beta)^{l-m}
\end{equation}
The summation on the left-hand side is trivial and yields
\begin{equation}
\int_0^{\pi}(\alpha\cos\theta+ i\beta\sin\theta)^{l} 
F_m^{l}(\theta) d\theta=\alpha^m(i\beta)^{l-m}.
\end{equation}
In the next step we change variables,
$\delta=(\alpha+\beta)/2$, $\gamma=(\alpha-\beta)/2$ and we have
\begin{equation}
\int_0^{\pi}(\delta e^{i\theta}+\gamma e^{-i\theta})^{l} 
F_m^{l}(\theta) d\theta=(\gamma+\delta)^m[i(\delta-\gamma)]^{l-m}.
\label{FFOUR}
\end{equation}
Now we set $\delta=1$, differentiate (\ref{FFOUR}) $k$-times with respect 
to $\gamma$ and then set $\gamma=0$. After little algebra we arrive at
\begin{eqnarray}
&&\int_0^{\pi} e^{i(l-2k)\theta} 
F_m^{l}(\theta) d\theta=
\nonumber \\
&&i^{l-m}\frac{(l-k)!}{l!}
\left.\frac{d^k}{d\gamma^k}
\left[(1+\gamma)^m(1-\gamma)^{l-m}\right]\right|_{\gamma=0}.
\label{FFIVE}
\end{eqnarray}
Now we assume that the function $F_m^l$ can be written in terms of
 finite Fourier series,
 \begin{equation}
 F_m^l(\theta)=\sum_{n=0}^l e^{i(l-2n)\theta} E_{mn}^l,
 \label{FSUM}
 \end{equation}
 and insert this expansion into (\ref{FFIVE}). After
 integration on the left-hand side and differentiation on the right-hand side
 of (\ref{FFIVE}) we find
 \begin{equation}
 E_{mn}^l=\frac{i^{l-m}}{\pi}{l \choose n }^{-1}\sum_{j=0}^{m}
 {m \choose j} {l-m \choose n-j} (-1)^{n-j},
 \end{equation}
 and we have derived the formulas (\ref{FORT}) and (\ref{EFORT}).

\end{multicols}
\end{document}